\documentclass[11pt]{article}

\usepackage{graphicx}

\begin{document}

{\LARGE \bf On the orbit of the LARES satellite}\\
\vspace{28pt}

\begin{centering}
I. Ciufolini\\
\end{centering}
\vspace{.25in}

Dipartimento di Ingegneria dell'Innovazione, Universit\`{a} di
Lecce, Via Monteroni, 73100 Lecce, Italy\\

{\Large \bf Abstract}

This paper is motivated by
the recent possibility to find an inexpensive launching vehicle for the
LARES satellite, however at an altitude much lower than originally planned
for the LAGEOS III/LARES satellite.

We present here a preliminary error analysis corresponding to a
lower, quasi-polar, orbit, in particular we analyze the effect on the LARES
node of the Earth's static gravitational field, and in particular
of the Earth's even zonal
harmonics, the effect of the time dependent Earth's gravitational field,
and in particular of the $K_1$ tide, and the effect of particle drag.

\section{Introduction}

In 1984 we proposed the use of the nodes of two laser ranged satellites
of LAGEOS type to measure the Lense-Thirring effect \cite{ciu84,ciu86}.
We proposed to orbit a laser-ranged
satellite of LAGEOS-type (called LAGEOS III and later on LARES),
with an inclination supplementary to the one
of LAGEOS (launched in 1976) in order to cancel out all the secular effects
on the nodes of the two laser-ranged satellites due to the deviations
of the Earth's gravitational field from spherical symmetry and in particular
due to the Earth's even zonal harmonics. All the other orbital parameters of
LARES/LAGEOS III were proposed to be equal to the ones of LAGEOS,
in particular the semimajor axis was proposed to be approximately
equal to 12270 km.
The mass of LAGEOS and of the proposed LAGEOS III satellite
was about 400 kg.

Unfortunately, even though such an orbit of LARES would have allowed
a complete cancellation of the static
Earth's spherical harmonics secular effects in order to measure the much
smaller Lense-Thirring effect, the weight of the proposed LARES satellite
of about 400 kg and especially the high altitude of its orbit implied
an expensive launching vehicle. For this reason a LARES satellite
of only about 100 kg of weight was later designed \cite{ciup}, nevertheless
the high altitude of LARES was still somehow expensive to achieve.

Nevertheless, three new factors have changed the need of
such a high altitude orbit for LARES: (a) the idea to use the
nodes of N laser-ranged satellites to measure the Lense-Thirring
effect and to cancel the uncertainty due to the first N-1 even zonal
harmonics, (b) the launch of the GRACE satellite in 2002 and the publication
of a new generation of very accurate Earth's gravity field models using the
GRACE observations and (c) the possibility to launch the LARES satellite
using an inexpensive launcher, however at a much lower altitude than
originally planned.

(a) The idea to use the nodes of N satellites of LAGEOS type to cancel
the effect of the first N-1 Earth even zonal harmonics and to
measure the Lense-Thirring effect was published in 1989 \cite{ciu89}
(see also the 1995 book \cite{ciuwhe}, on page 336)
as a possible alternative to the concept of the
supplementary inclination satellites. This technique and
in particular the idea to use the two nodes of the satellites
LAGEOS  and LAGEOS 2, together with the perigee of LAGEOS 2, was
described in details, together with the corresponding formula, in 1996
\cite{ciu96}.
In the 1989 paper (see also \cite{ciuwhe}), in order to measure the
Lense-Thirring effect and to cancel the even zonal harmonics
uncertainties, it was proposed: ``For $J_2$, this
corresponds from formula (3.2), to an uncertainty in the nodal
precession of 450 milliarcsec/year, and similarly for higher
$J_{2n}$ coefficients. Therefore the uncertainty in $\dot
\Omega^{Lageos}$ is more than ten times larger than the
Lense-Thirring precession. A solution would be to orbit several
high-altitude, laser-ranged satellites, similar to LAGEOS, to
measure $J_2, J_4, J_6$ etc, and one satellite to measure $\dot
\Omega_{Lense-Thirring}$''. At that time the error due to
the even zonal harmonics was quite large due to the much less
accurate Earth gravity models (available at that time) and the
LAGEOS 2 satellite was not yet launched (it was launched in 1992).

This technique to use N observables to cancel the effect
of the first N-1 even zonal harmonics was explicitly
described in \cite{ciu96} (see also the explicit calculations
about the use of the nodes of N laser-ranged satellites in
\cite{pet}) and
led to the publication
of the detection of the Lense-Thirring effect using the LAGEOS
and LAGEOS 2 satellites and the gravity field model
EGM96 \cite{ciup98,ciuetal97} (using the nodes of LAGEOS
and LAGEOS 2 and the perigee of LAGEOS 2 in order
to cancel the error in the
first two even zonal harmonics by using three ''observables'',
including the perigee, which however introduced relatively large
errors due to its unmodelled non-gravitational perturbations)
and to the 2004 measurement \cite{ciupav,ciupavper}
(with accuracy of the order of about 10 $\%$)
of the Lense-Thirring effect using the LAGEOS
satellites and the accurate Earth's gravity field model EIGEN-GRACE02S,
published by the GeoForschungsZentrun of Potsdam (GFZ), Germany,
using the data of the GRACE satellite.
In this 2004 paper we used the same technique of the 1998 paper but
we did not use the perigee of LAGEOS II (that the authors
of \cite{ciupav} tried for a long time to avoid since the publishing
of their 1997-1998 papers) thanks to the new generation of GRACE Earth's
gravity models. The 2004-measurement is just
the case of N=2 described in the abovementioned 1989 paper
and it uses the nodes of the two
laser ranged satellites LAGEOS and LAGEOS 2 in order
to cancel the effect
of the first even zonal harmonic coefficient $J_2$ of Earth and to
measure the Lense-Thirring effect (the explicit expression
of this combination was also given in \cite{ciu02}.

(b) The 2004 accurate measurement of the Lense-Thirring effect was
possible thanks to the launch of the GRACE satellite and the
publication of its accurate gravity field models by GFZ \cite{rei}
and Center for Space Research (CSR) of the University
of Texas at Austin. The use of the GRACE-derived gravitational models, when
available, to measure the Lense-Thirring effect with accuracy of a
few percent was, since almost a decade ago, a well know
possibility to all the researchers in this field and was presented
(and published) at several meetings by Pavlis \cite{ecp} and by
Ries et al. (see, e.g., \cite{rie}).

(c) In 2004, A. Paolozzi of ``Scuola d'Ingegneria Aerospaziale'' 
of the University of Rome ``La Sapienza'' discovered the possibility
to use an inexpensive launcher to orbit
LARES \cite{pao04}. However, this inexpensive
launch for LARES should be at a much lower altitude than the originally
planned satellite at 12270 km and
should be in a nearly polar orbit. The altitude achievable with
this launch should be between about 1000 km and 2000 km.
In 2005, J. Ries \cite{rie05} informed us that CSR had done some simulations supporting
this possibility of a lower orbit laser-ranged satellite. This was
also suggested to us by P. Bender \cite{ben05}.

In the following we shall investigate on the possible orbit
of the LARES satellite in relation to the error in the measurement
of the Lense-Thirring effect.

\section{Preliminary error analysis for a quasi-polar laser-ranged
satellite at an altitude of about 1500 km}

The simplest conceivable orbit in order to cancel the effect
of all the even zonal harmonics on the node of a satellite would
be a polar orbit, indeed for such an orbit the effect of the even zonal
harmonics on the satellite node would be zero and, however, the node of the
satellite would be still perturbed by the Earth's gravitomagnetic
field, i.e., would be affected by the Lense-Thirring effect.

Unfortunately, as pointed out in the 1989 LAGEOS III NASA/ASI
study \cite{NASA-ASI}) and explicitely calculated by Peterson
(1997) (chapter 5 of \cite{pet}) the uncertainty in the $K_1$ tide
(tesseral, $m=1$, tide) would make such an orbit unsuitable for
the Lense-Thirring measurement. Indeed, a polar satellite would
have a secular precession of its node whose uncertainty would
introduce a large error in the Lense-Thirring measurement. In
addition, it would be quite demanding to launch LARES with the
requirement of a small orbital injection error from a polar orbit 
(even at a lower altitude than the one planned for LAGEOS III, 
for a measurement error of about 1 $\%$
due to the uncertainties in the static Earth gravity field,
the deviation from a polar orbit 
should be less than about 0.1 degrees).

Nevertheless, a quasi-polar orbit would have a nodal
precession, due to its departure from 90 degrees of
inclination, and thus one could simply fit for the effect
of the $K_1$ tide using a periodical signal exactly at the
nodal frequency. Such frequency (with the period of the LAGEOS
satellites node)
is indeed observed in the LAGEOS 1 and LAGEOS 2 analyses already
mentioned \cite{ciup98,ciupav}
and is the largest periodical amplitude observed in the combined residuals.

If we assume that the LARES orbit would have an altitude of 1500 km, then,
by imposing for example that the minimum period of observation in order to
measure the Lense-Thirring effect should not be
longer than three years, we have that the LARES inclination
should be less than or equal to 86 degrees or larger than
or equal to 94 degrees.

In regard to the effect of the static even zonal harmonics,
by using the technique explained in \cite{ciu89,ciu96} and
by using the nodes of the satellites LARES, LAGEOS and LAGEOS 2,
we would be able to cancel the uncertainties due
to the first two even zonal harmonics, $C_{20}$ and $C_{40}$, and our
measurement will only be affected by the
uncertainties due to the even zonal harmonics with degree strictly
higher than 4.

By solving the system of the three equations for the nodal
precessions of LAGEOS, LAGEOS I and LARES in the three unknowns,
$J_2$, $J_4$ and Lense-Thirring effect, we have a combination of
three observables (the three nodal rates) which determines the
Lense-Thirring effect independently of any uncertainty $\delta
C_{20}$ and $\delta C_{40}$ in the first two even zonal harmonics.
This same technique was applied in \cite{ciup98} using the nodes
of LAGEOS and LAGEOS 2 and the perigee of LAGEOS 2 and in
\cite{ciupav} using the nodes of LAGEOS and LAGEOS 2 only.

It turns out that some values of the inclination of LARES
would minimize the error in the measurement of the
Lense-Thirring effect since they would minimize the error due
to the uncertainty in the largest (not cancelled using the
combination of the three observables) even zonal harmonic $C_{60}$.

In figure 1 we have plotted the error in the measurement
of the Lense-Thirring effect, using LARES, LAGEOS
and LAGEOS 2, as a function of the
inclination and of the semimajor axis. The range of the altitude
of LARES is between 1000 km and 2000 km and of the inclination between
0 and 360 degrees, of course if LARES would be launched
in a nearly polar orbit the
use of the LAGEOS and LAGEOS 2 satellites would not be
anymore useful in order to reduce the error budget (and would indeed
only introduce an additional error), since the
effect of the even zonal harmonics on the node of LARES would
be nearly zero, however, as previously remarked, 
the measurement of the Lense-Thirring effect using
a polar orbit would be substantially affected by the uncertainty in the $K_1$ tide.

In figure 2 we have plotted the error in the measurement
of the Lense-Thirring effect as a function of the
inclination by assuming an altitude of LARES
of 1500 km, i.e., a LARES semimajor axis of about 7870 km.

From Figure 2 we can see that any inclination from 60 degrees to
86 degrees and from 94 to 120 degrees would be suitable for a
measurement of the Lense-Thirring effect with accuracy of a few 
percent. An inclination of LARES of about 110
degrees or 70 degrees would minimize the error. In deriving this
result, we have assumed: (a) zero eccentricity for the LARES orbit,
(b) we have only considered the effect of the first 5 even zonal
harmonics \cite{tau04}: $C_{20}$, $C_{40}$, $C_{60}$, $C_{80}$ and
$C_{10 \, 0}$ and (b) we have considered the uncertainties in the
spherical harmonics $C_{60}$, $C_{80}$ and $C_{10 \, 0}$ to be
equal to those of the EIGEN-GRACE02S Earth's gravity model
\cite{rei}, i.e., we have assumed $\delta C_{60}=0.2049 \cdot 10^{-11}$
$\delta C_{80}=0.1479 \cdot 10^{-11}$ $\delta C_{10 \,
0}=0.2101 \cdot 10^{-11}$. Nevertheless, by including higher degree even
zonal harmonics, by considering an eccentricity different from
zero, e.g., equal to 0.01, and by considering an error a priori
equal for these three spherical harmonics, i.e. $\delta
C_{60}=\delta C_{80}=\delta C_{10 \, 0}=0.2 \cdot 10^{-11}$, Figures 1
and 2 would not appreciably change and our results would still
remain valid. This uncertainty, $0.2 \cdot 10^{-11}$, is nearly 
the uncertainty in the even zonal
coefficients of the EIGEN-GRACE02S model (used in \cite{ciupav})
and given above); indeed, even though the real error in these
coefficients would probably be about two times larger than these
published values, these uncertainties refer to a preliminary 2004
model and by the time of the launch of LARES and of its data
analysis (about 2008-2011), Earth's gravity field models much more
accurate based on much longer data set of GRACE observations would
be available.

In regard to the other orbital perturbations that affect the
LARES experiment we briefly discuss here the tidal effects,
particle drag and thermal drag; for a detailed total error budget
we refer to \cite{ciu89,NASA-ASI,ciuetal07}. In regard to the
orbital perturbations on the LARES experiment due to the time
dependent Earth's gravity field, we observe that the largest tidal
signals are due to the zonal tides with $l=2$ and $m=0$, due to
the Moon node, and to the $K_1$ tide with $l=2$ and $m=1$
(tesseral tide). However, the medium and long period zonal tides
($l=2$ and $m=0$) will be cancelled using the combination of the
three nodes together with the static $C_{20}$ uncertainty (also
the uncertainty in the time-dependent secular variations $\dot
C_{20}$, $\dot C_{40}$ will be cancelled using this combination of
three observables). Furthermore, the tesseral tide $K_1$ will be
fitted for over a period equal to the LARES nodal period as
explained above (see \cite{NASA-ASI} and chapter 5 of \cite{pet}) and this tide would
then introduce a small uncertainty in our combination. In regard
to the non-gravitational orbital perturbations, we observe here
that the unmodelled thermal drag perturbations on the LARES orbit
would be reduced thanks to the accurate measurements of the
thermal properties of the LARES satellite and of its
retro-flectors that are performed by the group 
of S. Dell'Agnello at the Laboratori Nazionali di
Frascati of INFN \cite{infn06}. We finally point out that the
neutral and charged particle drag on the LARES node at an altitude
of about 1500 km would be a negligible effect for an orbit with
very small eccentricity, even by assuming that the exosphere would
be co-rotating with Earth at 1500 km of altitude. Indeed, as
calculated in \cite{ciu89} for the LAGEOS III satellite, in the
case of zero orbital eccentricity $e=0$ the total drag effect on
the LARES node would be zero; indeed the nodal rate of a satellite due to
particle drag is a function of $sin \; \nu \cdot cos  \; \nu$
($\nu$ is the true anomaly) and the total nodal shift is then zero
over one orbit; in the case of a small orbital eccentricity, the
total shift would be proportional to the eccentricity and it would
still be a small effect as calculated in \cite{ciu89}.

\begin{figure}
\begin{center}
\includegraphics[scale=.5]{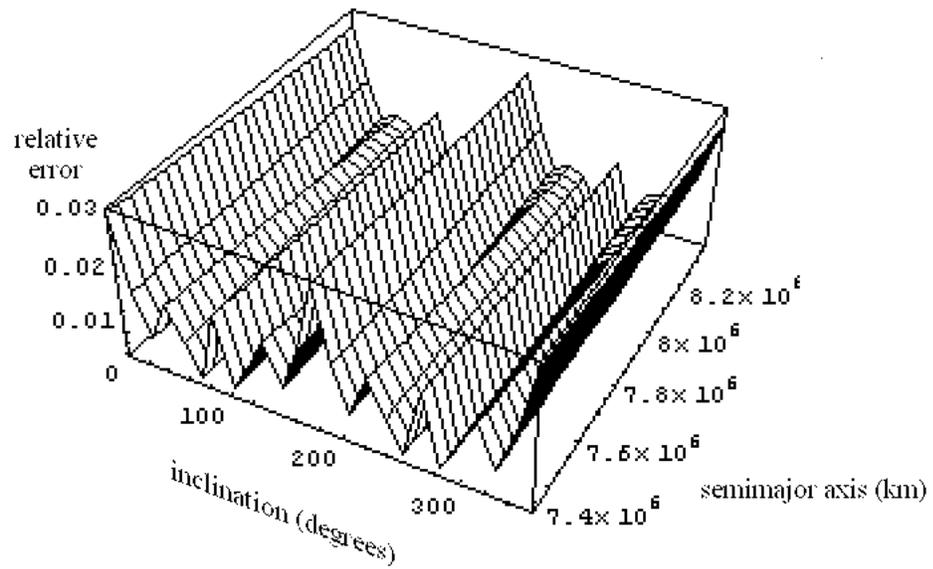}
\caption{Uncertainty in the measurement
of the Lense-Thirring effect, due to the even zonal
harmonics uncertainties, as a function of the
inclination and of the semimajor axis of LARES, using LARES, LAGEOS
and LAGEOS 2. The range of the altitude of LARES is
between 1000 km and 2000 km and of the inclination between
0 and 360 degrees}
\end{center}
\end{figure}

\begin{figure}
\begin{center}
\includegraphics[scale=.5]{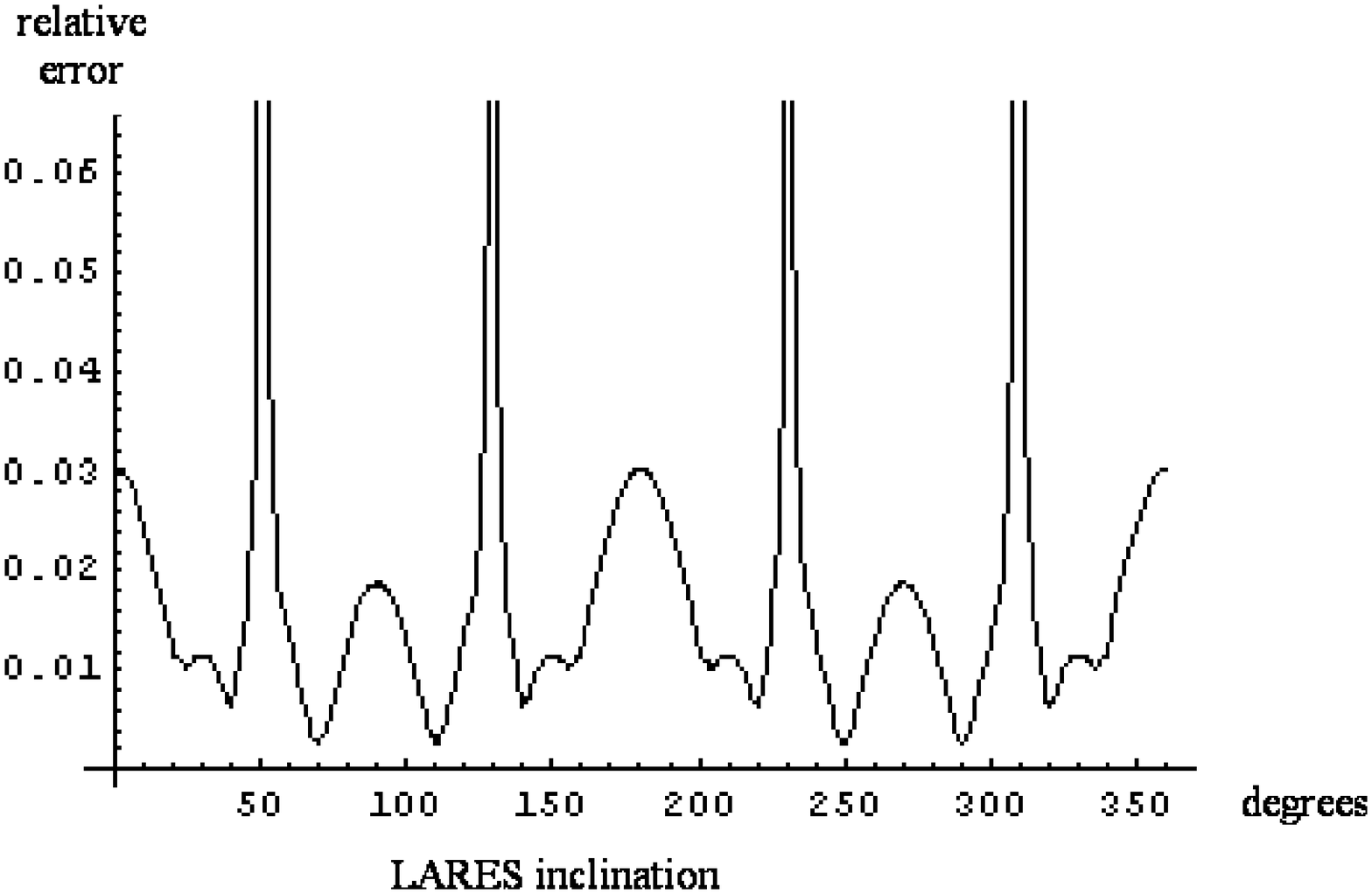}
\caption{Uncertainty in the measurement
of the Lense-Thirring effect, due to the even zonal
harmonics uncertainties, as a function of the
inclination of LARES, using LARES, LAGEOS
and LAGEOS 2. The altitude of LARES is here
1500 km and the range of the inclination between
0 and 360 degrees}

\end{center}
\end{figure}

\section{Conclusion}

A nearly polar orbit for LARES at an altitude of about 1500 km would
be suitable for a measurement of the Lense-Thirring effect with accuracy
of a few percent. Some values of the inclination of LARES
would minimize the measurement error induced by the uncertainties
in the even zonal harmonics. An inclination off the polar one by
about 4 degrees would allow
the average and the fit of the $K_1$ tidal uncertainty over a period of about 3 years.

\end{document}